\documentclass[a4paper,10pt,twoside]{just}

\usepackage{multicol}
\usepackage{graphicx}
\usepackage{booktabs}
\usepackage{amssymb,bm,mathrsfs,bbm,amscd}
\usepackage[tbtags]{amsmath}
\usepackage{lastpage}
\usepackage{CJK}
\usepackage{xspace}

\newcommand{\sw}{$\sin^2\theta_W$\xspace}

\begin{document}
\begin{CJK*}{GBK}{song}

\fancyhead[c]{\small  10th International Workshop on e+e- collisions from Phi to Psi (PHIPSI15)}
 \fancyfoot[C]{\small PHIPSI15-\thepage}

\footnotetext[0]{Received 12 Nov. 2015}

\title{Measuring the weak mixing angle with the P2 experiment at MESA\thanks{Supported by 
the cluster of excellence PRISMA and the Collaborative Research Center 1044, both
funded through the Deutsche Forschungsgemeinschaft (DFG)
}}

\author{%
      Niklaus~Berger$^{1;1)}$\email{niberger@uni-mainz.de},
			Kurt~Aulenbacher$^{1,2}$,
			Sebastian~Baunack$^{1}$,
			Dominik~Becker$^{1}$,
			J\"urgen~Diefenbach$^{1}$,\\
			Michael~Gericke$^{3}$,
			Kathrin~Gerz$^{1}$,
			Ruth~Herbertz$^{1}$,
			Krishna~Kumar$^{4}$
			Frank~Maas$^{1,2}$,\\
			Matthias~Molitor$^1$,
			David Rodr\'iguez Pi\~neiro$^{1,2}$,
			Iurii~Sorokin$^1$,
			Paul~Souder$^5$,
			Hubert~Spiesberger$^6$,\\
			Alexey~Tyukin$^1$,
			Valery~Tyukin$^1$
			and
			Marco~Zimmermann$^1$
}
\maketitle

\address{%
$^1$ PRISMA Cluster of Excellence and Institute of Nuclear Physics, Johannes Gutenberg University, Mainz, Germany\\
$^2$ Helmholtz Institue Mainz, Germany\\
$^3$ Department of Physics and Astronomy, University of Manitoba, Winnipeg, Canada\\
$^4$ Department of Physics and Astronomy, Stony Brook University, Stony Brook, USA\\
$^5$ Physics Department, Syracuse University, Syracuse, USA\\
$^6$ PRISMA Cluster of Excellence and Institute of Physics, Johannes Gutenberg University, Mainz, Germany\\

}

\begin{abstract}
The P2 experiment in Mainz aims to measure the weak mixing angle \sw in electron-
proton scattering to a precision of 0.13\%. In order to suppress uncertainties
due to proton structure and contributions from box graphs, both a low average
momentum transfer $Q^2$ of $4.5\cdot 10^{-3}$~GeV$^2/c^2$ and a low beam energy of 
155~MeV are chosen. In order to collect the enormous statistics required for this 
measurement, the new Mainz Energy Recovery Superconducting Accelerator (MESA) is
being constructed. These proceedings describe the motivation for the measurement,
the experimental and accelerator challenges and how we plan to tackle them.
\end{abstract}



\begin{multicols}{2}

\section{Introduction}

The weak mixing angle \sw is one of the fundamental parameters of the Standard
Model (SM) of elementary particle physics. It has been measured with great
precision at the $Z$ resonance \cite{ALEPH:2005ab}, where the determinations from
LEP and SLD are marginally consistent. Due to quantum corrections, the effective
weak mixing angle is a scale dependent quantity and measurements at 
different scales become important both for testing the SM and searching for
effects of new physics beyond the SM in the running \cite{Erler:2003yk, Erler:2004cx}.

Measurements at lower scales were obtained in neutrino nucleon scattering \cite{Zeller:2001hh}, 
deep inelastic electron scattering \cite{Aktas:2005iv}, parity 
violating electron scattering on electrons \cite{Anthony:2005pm} and protons 
\cite{Androic:2013rhu} and atomic parity
violation in Caesium \cite{Wood:1997zq}.
These measurements were sufficient to establish the running of
\sw, more precision is however required for a stringent test of the SM and searches
for new physics.
The final result from the Qweak experiment is eagerly awaited and should improve
on the published result \cite{Androic:2013rhu} by a factor of
three to four. More precise determinations require new experimental approaches,
such as the proposed M\o ller \cite{Benesch:2014bas} and deep inelastic (SOLID, \cite{Chen:2014psa}) scattering experiments 
at JLAB and the P2 experiment in Mainz, which will be described in the following.
An overview of current and planned experiments together with the theory prediction \cite{Erler:2004in}
for the running of \sw is shown in Fig.~\ref{fig1}.

A precise determination of the weak mixing angle at low scales is sensitive to contributions of new
physics  
\begin{center}
\includegraphics[width=0.5 \textwidth]{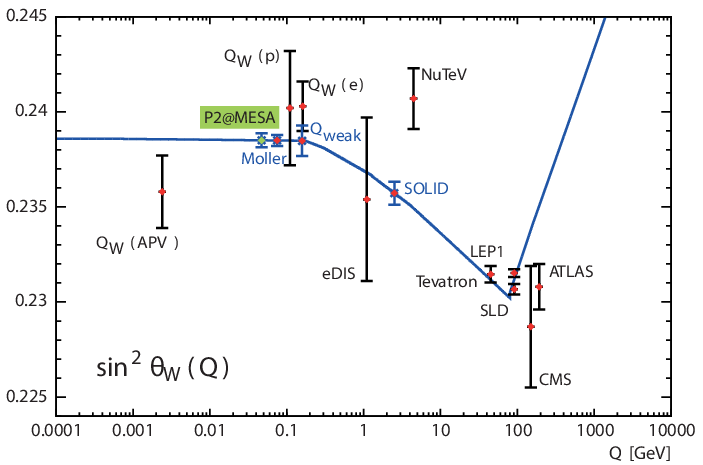}
\figcaption{\label{fig1} Scale dependence of \sw together with completed (black
error bars) and planned (blue error bars, value chosen to coincide with theory)
experimental measurements.}
\end{center}
beyond the SM which can change the running of \sw 
via contributions of
new gauge bosons, additional fermions, 
mixing terms \cite{Davoudiasl:2014kua} 
or the exchange of very heavy particles
 which can be parametrized as four-fermion contact interactions \cite{Erler:2003yk}.
 In the
last case, P2 will be sensitive to scales up to 49~TeV, comparable to the experiments
at the large hadron collider after collecting 300~fb$^{-1}$ of integrated luminosity.

\section{Requirements}

The P2 experiment aims to determine \sw with a precision of 0.13\% by performing
a measurement of the parity violating asymmetry $A_{PV}$ in electron-proton
scattering. This asymmetry between the cross-sections for left- and righthanded electrons
$\sigma_L$ and $\sigma_R$ is determined by the weak charge of the proton $Q_W$:
\begin{eqnarray}
A_{PV} = \frac{\sigma_L-\sigma_R}{\sigma_L+\sigma_R} = \frac{G_F Q^2}{4 \sqrt{2}\pi\alpha}(Q_W + F(Q^2)),
\end{eqnarray}
where $G_F$ is the Fermi constant, $\alpha$ the fine structure constant and $Q^2$
the squared four-moment transfer, setting the scale. Contributions stemming from
the fact that the proton is not a point-like particle are collected in $F(Q^2)$
and are small for low values of $Q^2$, thus motivating an experiment at low
momentum transfer. 
Further hadronic uncertainties come from box graphs such as the one shown in Fig.~\ref{fig2}.
These uncertainties have a weak dependence on the momentum transfer, do however
increase steeply with rising center-of-mass energy \cite{Gorchtein:2011mz, Gorchtein:2015naa}, 
favouring a low beam energy.

The weak mixing angle is related to $Q_W$ via
\begin{eqnarray}
Q_W = 1 - 4 \sin^2\theta_W,
\end{eqnarray}
which implies by propagation of uncertainty that a 0.13\% measurement of \sw 
requires a 1.5\% measurement of $Q_W$, which also corresponds to the target uncertainty
in the asymmetry. Due to the small weak charge of the proton and the small $Q^2$,
the expected asymmetry is only 33~ppb, thus requiring a measurement with 0.44~ppb
precision. The statistical uncertainty scales with the number of scattered electrons
as $\frac{1}{\sqrt{N}}$, which in turn requires the observation of $\mathcal{O}(10^{18})$
electrons. For sociological reasons, the total measurement time is limited to 10'000~hours,
which requires observing $\mathcal{O}(10^{11})$ signal electrons per second. 

These very high rates can be achieved by directing a 150~$\mu$A electron beam
onto a 60~cm long liquid hydrogen target, producing a luminosity of 
$2.4\cdot 10^{39}$~s$^{-1}$cm$^{-2}$.

The aim of determining \sw with a precision of 0.13\% is thus extremely challenging
for the accelerator and detector systems. The following sections outline how the
MESA accelerator, the polarisation measurement and the P2 experiment intend to 
tackle these challenges.

\begin{center}
\includegraphics[width=0.24 \textwidth]{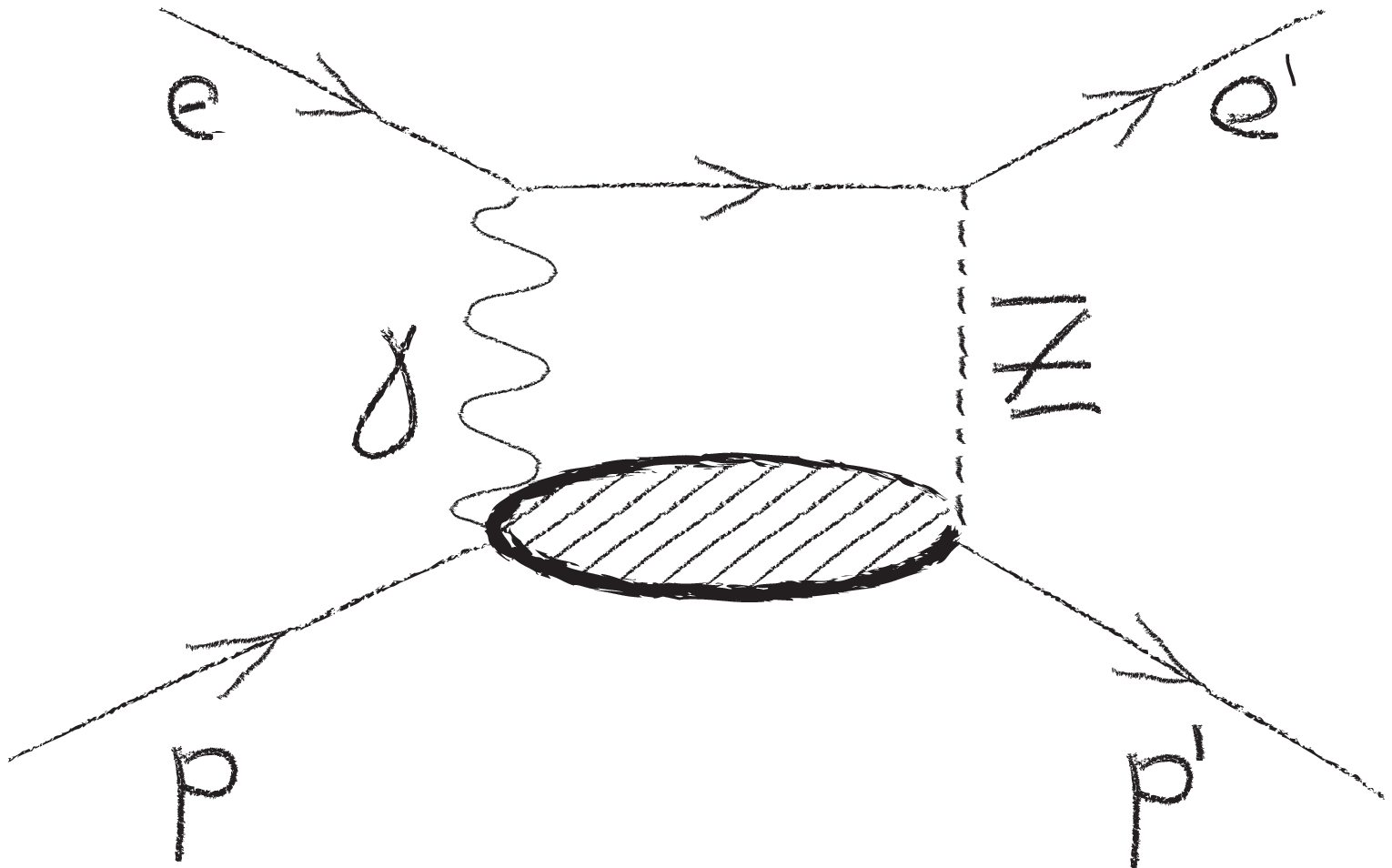}
\figcaption{\label{fig2} $\gamma-Z$ box graph for electron-proton scattering.
The hadronic uncertainty stems from the possible excited states of the proton
indicated by the shaded blob.}
\end{center}

\section{The MESA accelerator}

In order to accommodate the very long running time and demanding stability 
requirements of the P2 experiment, a new accelerator, the Mainz Energy-Recovery
Superconducting Accelerator (MESA, \cite{Aulenbacher:2013xla}) is being built.

With a maximum extracted beam energy of 155~MeV, MESA is small enough to fit into
the existing halls that have become available with the completion of the A4 parity
violating electron scattering program at the Mainz Microtron MAMI. 
P2 and the MAGIX spectrometer
(see the contribution of A.~Denig to this conference for details on the MESA
program beyond the P2 experiment) will be housed in a new hall as part of the
recently funded centre for fundamental physics. Fig.~\ref{fig3} shows the overall
layout of accelerator and experiments.

P2 requires a highly polarized ($>$ 85\%), high intensity (150~$\mu$A) beam of
155~MeV electrons with excellent availabiliy ($>$ 4000~h/year). The beam helicity
will be flipped several thousand times a second. The main challenge is to
reduce any helicity correlated changes in beam intensity, energy, position and
angle to less than 0.1~ppb. Here we can profit from the extensive experience in
beam stabilization gained at the Mainz Microtron MAMI. Table~\ref{tab1} compares
the values for helicity correlated beam fluctuations achieved at MAMI with the
requirements for P2 at MESA. Whilst the energy stability already fulfills the
demands, improvements of one to two orders of magnitude have to be achieved for 
position, angle and intensity; new digital feedback electronics for beam stabilization
are currently being designed and tested at MAMI.
 
\begin{center}
\tabcaption{ \label{tab1}  Helicity correlated beam fluctuations.}
\footnotesize
\begin{tabular*}{80mm}{c@{\extracolsep{\fill}}rrr}
\toprule 
Beam      & Achieved & Contribution & Required\\
Quantity  & at MAMI  & to $\delta(A_{PV})$ & for MESA \\
\hline
Energy 		& 0.04~eV & $<$ 0.1~ppb & fulfilled \\
Position  & 3~nm    &     5~ppb & 0.13~nm \\
Angle     & 0.5~nrad & 		3~ppb & 0.06~nrad \\
Intensity & 14~ppb  &     4~ppb & 0.36~ppb\\
\bottomrule
\end{tabular*}
\end{center}

The MESA lattice design is finalized, the superconducting RF cavities have been
ordered and civil construction on the new hall will start 2016. We plan to start
installing the accelerator in 2018 and have beam available for P2 before 2020.

\end{multicols}
\begin{center}
\includegraphics[width=12.0cm]{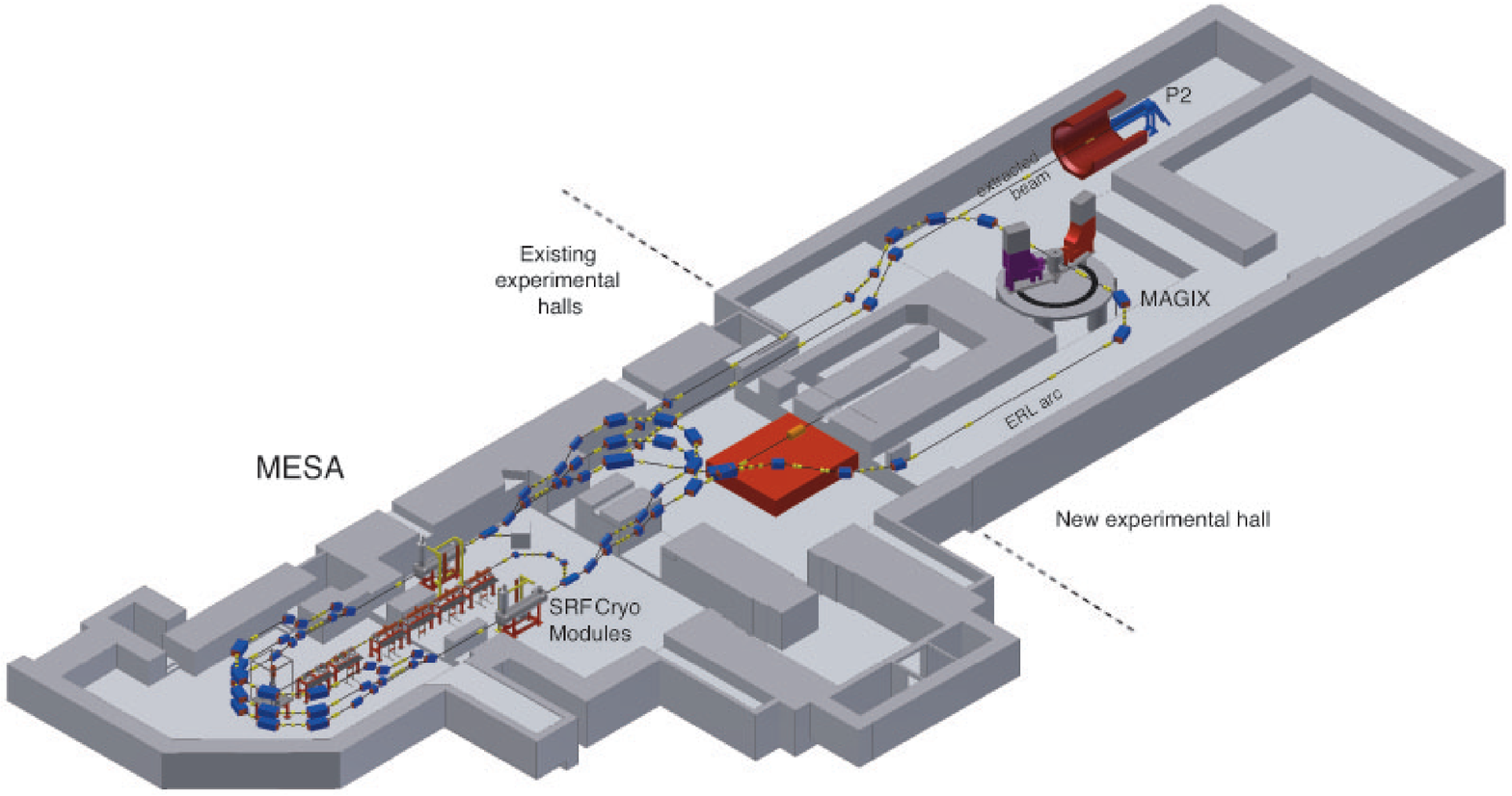}
\vspace{2mm}
\figcaption{\label{fig3} Layout of the MESA accelerator and experiments, indicating
existing and new halls at the institute of nuclear physics in Mainz.}
\end{center}
\begin{multicols}{2}

\section{Polarimetry}

P2 requires a knowledge of the beam polarisation of better than 0.5\%. We aim to
achieve this precision via two paths \cite{{Aulenbacher:2012um}},
namely an invasive double Mott polarimeter
at the electron source and a Hydro-M\o ller polarimeter, which can be operated at
the same time as and is placed right in front of the main experiment.

\subsection{Double Mott polarimeter}

The asymmetry of Mott scattering in thin foils can be used to determine the beam
polarization, it however requires a precise knowledge of the analyzing power of 
the scattering foils which introduces a large uncertainty into the measurement. 
Double Mott scattering \cite{Gellerich1991} in two foils 
allows to determine the effective analyzing power within the setup, thus reducing
the associated uncertainty which makes it a suitable choice for precise source 
polarimetry at MESA. A prototype of the double-Mott
polarimeter is currently tested with the MESA source prototype in operation in Mainz.

\subsection{Hydro-M\o ller polarimeter}

We plan to determine the beam polarization at the final energy with a hydro-M\o ller
polarimeter \cite{{Aguar2013}} right in front of the main experiment. Here the 
asymmetry in M\o ller 
scattering of the beam electrons with the electrons in fully polarized atomic 
hydrogen is used. The hydrogen is polarized using a 7-8~T solenoid magnet. In 
order to avoid hydrogen recombination, the gas is kept at cryogenic temperatures 
and the walls of the vessel are coated with superfluid helium. Operating this
cryogenic setup with a high intensity electron beam passing through the center is
certainly challenging. The cryostat/magnet for this setup is currently under construction.

\section{The P2 experiment}

For a given electron beam energy, the scattering angle $\vartheta$ determines
the momentum transfer $Q^2$. At low $Q^2$, the uncertainty of the asymmetry
measurement is dominated by statistics and helicity correlated beam fluctuations.
At large $Q^2$, uncertainties in the proton form factors become dominant. For
our setup, the best accuracy can be reached with a central scattering angle of 35$^\circ$
at an angular acceptance of 20$^\circ$. 

\begin{center}
\includegraphics[width=0.4 \textwidth]{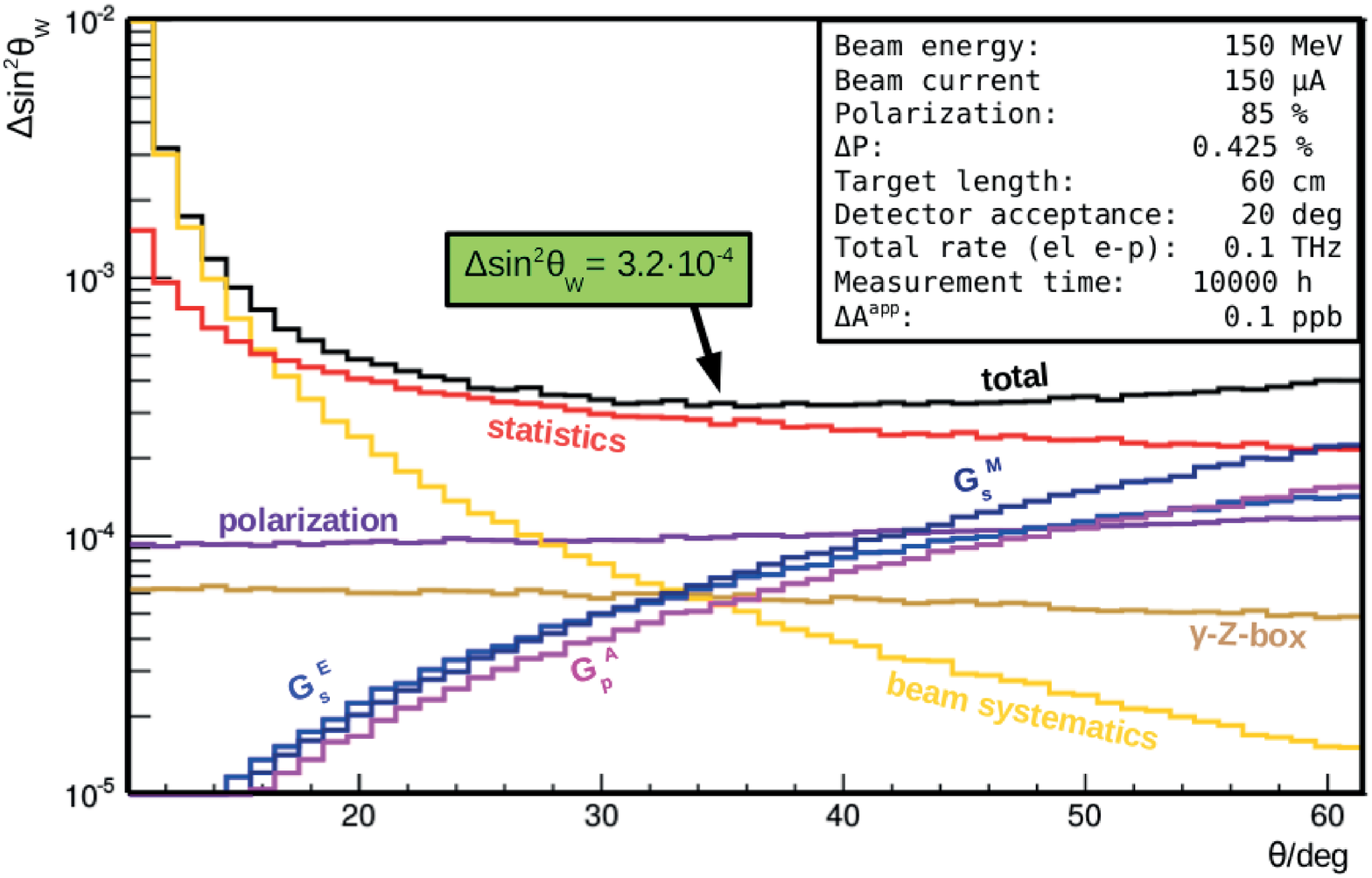}
\figcaption{\label{fig4} Contributions to the uncertainty of the \sw measurement
at fixed beam energy, intensity and run time in dependence of the central
scattering angle. $G_E^S$, $G_M^S$ and $G_P^A$ refer to the uncertainties stemming 
from the electric and magnetic strange form factor and the axial form factor of 
the proton respectively.}
\end{center}

As the very high intensity beam produces several thousand Bremsstrahlung
photons for every electron scattered into the angular range of interest as well as
a large amount of M\o ller scattered electrons with low transverse momentum, a
magnetic spectrometer is required in order to guide signal electrons to detectors
whilst at the same time shielding them from photon and M\o ller backgrounds.
The detectors in turn face the challenge of reliably detecting more than $100$~GHz
of scattered electrons. The following sections will describe the spectrometer design,
the development of integrating Cherenkov detectors as well as a pixellated tracking
detector for a precise determination of the momentum transfer $Q^2$.

\subsection{Spectrometer}

For the spectrometer design, the main choice is between a toroidal (as used in 
the QWeak experiment) and a solenoidal magnetic field. The advantages of 
a toroidal setup, such as zero field in the target region and easy access to
instrumentation are compromised by the fact that the coils are necessarily inside
of the spectrometer acceptance, typically leading to a loss of about half the 
signal electrons and consequently a doubling of the measurement time, which is
unrealistic in the context of P2.

We have thus decided to employ a solenoidal design and studied possible placements
of target, shielding and detectors for several existing solenoids, e.g.~from the
ZEUS experiment at HERA \cite{{Acerbi:1985yd}} or the FOPI experiment at GSI \cite{{Ritman:1995td}}. 
We have shown using both ray-tracing in the magnet field maps and full Geant4 based simulations that with a 
careful optimization of the shape and placement of lead shields, sufficient 
signal-to-background ratios can be achieved in the integrating detectors.
A possible view of the setup is shown in the rendering in Fig.~\ref{fig5}.

\begin{center}
\includegraphics[width=0.5 \textwidth]{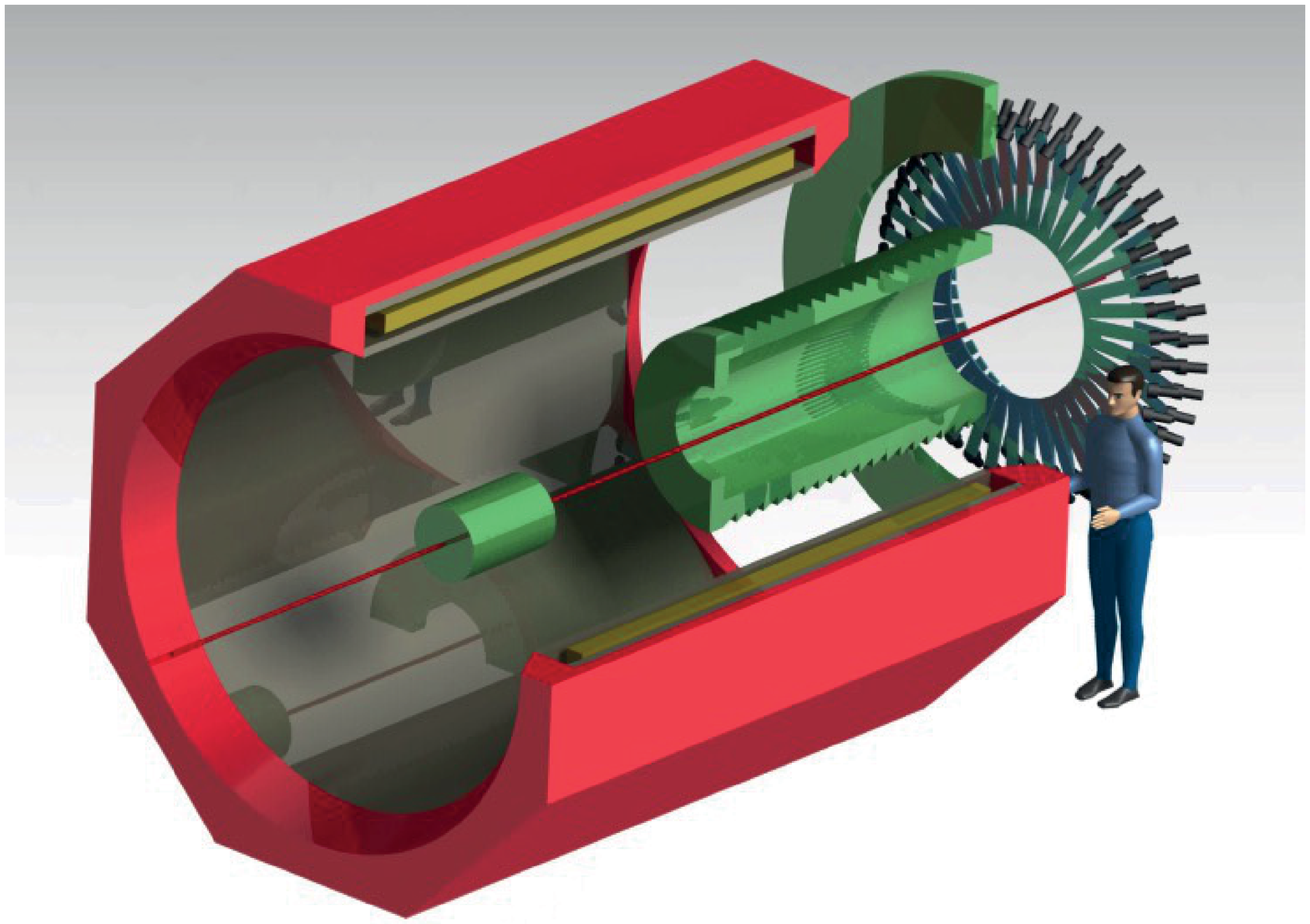}
\figcaption{\label{fig5} Rendering of a possible P2 setup, showing the solenoid
coil, target, lead shielding and integrating detectors as well as a hunky physicist 
for scale.}
\end{center}

\subsection{Integrating detectors}

Individually counting hundreds of GHz of electrons is extremely challenging, but 
not actually required for P2. Instead, we opt for an integrating measurement,
where the electrons produce Cherenkov light in bars of fused silica (quartz),
which is detected with photomultipliers operated at low gain. The current of these
photomultipliers is integrated over one helicity period and read out with high precision
(22~bit) analog-to-digital converters.

We are currently testsing different types and polishing finishes as well as wrappings
of quartz bars with the MAMI beam and have found both a performance sufficient for
P2 as well as an excellent agreement of the light yield at different incident angles 
with Geant4 based simulations. 

The switchable gain photomultiplier base is currently under development in Mainz;
for the integrating ADC a joint development with the M\o ller experiment is ongoing
at the University of Manitoba.

\subsection{Tracking detectors}

In order to determine the average momentum transfer $\langle Q^2\rangle$ of the scattered 
electrons creating signals in the integrating detectors, a tracking detector is
required. The high rates at MESA, the precision requirements and the low momenta
of the scattered electrons (making multiple coulomb scattering in the tracker 
material the dominating uncertainty in the momentum measurement) call for a fast, 
high granularity sensor with very little material. We choose to employ high-voltage
monolithic active pixel senors (HV-MAPS, \cite{Peric:2007zz, Peric:2013cka, 
Shrestha:2014oxa, Augustin:2015mqa, Peric:2015ska}) 
as the detector technology. These sensors, manufactured in a commercial CMOS
technology, apply a ``high'' voltage of around 90~V between deep $n$-wells
and the substrate, leading to very fast charge collection from a thin depletion layer.
The thin charge collection zone allows for thinning of the sensors to just 50~$\mu$m.
The sensor is segmented into 80 by 80~$\mu$m pixels.
The CMOS process used allows for integrating both analog and digital electronics
directly on the sensor; the output are zero suppressed hit addresses and timestamps
on a fast differential link.
In the development of the sensors for P2, we closely collaborate with the Mu3e,
ATLAS and Panda experiments.

\begin{center}
\includegraphics[width=0.5 \textwidth]{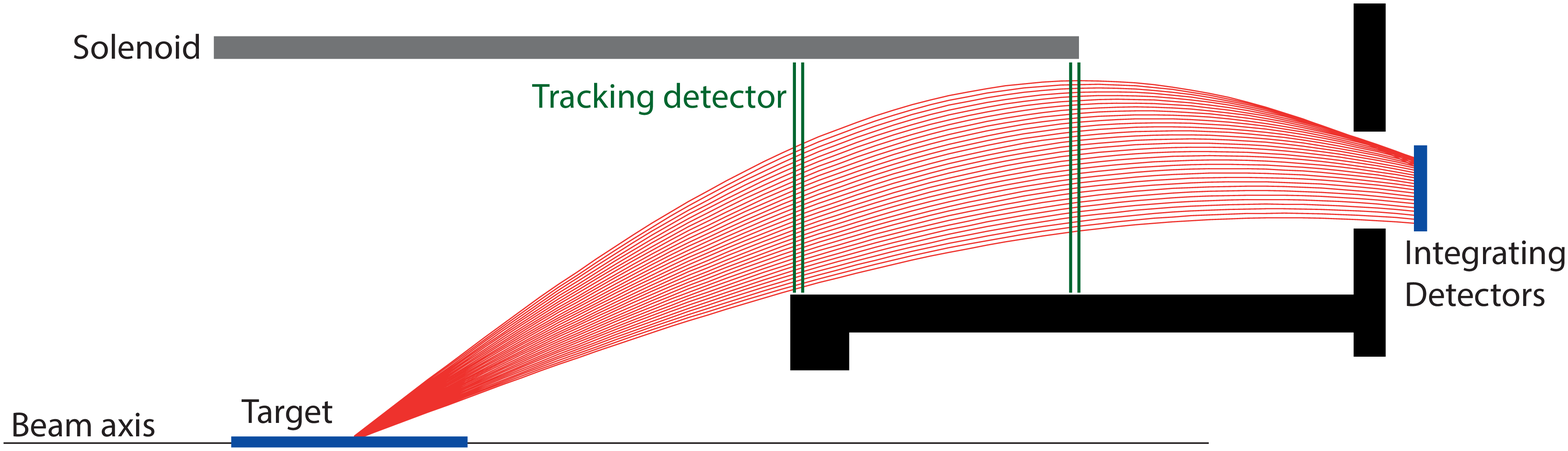}
\figcaption{\label{fig6} Schematic view of the P2 tracking detector, consisting
of four planes of HV-MAPS sensors.}
\end{center}

From these thin sensors, we plan to build a tracking detector with four planes,
see Fig.~\ref{fig6}. The arrangement in two double planes combines a good momentum 
and angular resolution in a multiple scattering dominated regime with ease of reconstruction
in a high multiplicity environment. We are currently studying tracking algorithms
that also perform well in the non-uniform field close to the edge of the magnet and 
are at the same time robust and fast enough to allow for on-line track finding and
fitting, possibly on highly parallel architectures such as graphics processing units
(GPUs).

\section{Summary and Outlook}

The P2 experiment aims to measure \sw at low momentum transfer with unprecedented
accuracy, which both improves the precision on one of the fundamental parameters
of the standard model and allows to search for new physics. The new MESA accelerator
in Mainz will provide a stable very high intensity electron beam combined with
precision polarimetry. P2 will measure the parity violating asymmetry in electron-
proton scattering using a solenoid spectrometer with integrating Cherenkov detectors
combined with a thin pixel tracker. Accelerator commissioning is scheduled to start
in 2018 and a first P2 data taking for 2020. Beyond the \sw measurement, P2 can
also be used to study parity violation with different target materials, giving
access e.g.~to neutron skins. The MESA accelerator also has a wide physics program
beyond P2, described elsewhere in these proceedings.

\vspace{5mm}
\centerline{\rule{80mm}{0.1pt}}

\end{multicols}

\clearpage

\end{CJK*}
\end{document}